\documentclass[twocolumn,prb]{revtex4}
\usepackage{amsmath}
\usepackage{amsfonts}
\usepackage{amssymb}
\usepackage{graphics}
\usepackage{float}
\usepackage{graphicx}
\usepackage{epstopdf}
\usepackage[compact]{titlesec}
\usepackage{natbib}
\usepackage{threeparttable}
\usepackage[titletoc,title]{appendix}
\usepackage{lipsum}

\begin{document}
\title{The tuning of light-matter coupling and dichroism in graphene for enhanced absorption: Implications for graphene-based optical absorption devices}
\author{Shaloo Rakheja$^{1}$}
\email{shaloo.rakheja@nyu.edu}
\email{parijats@bu.edu}
\author{Parijat Sengupta$^{2,3}$}
\affiliation{$^{1}$ Electrical and Computer Engineering, New York University, New York, NY, 11201 \\
$^{2}$ Photonics Center, Boston University, Boston, MA, 02215 \\
$^{3}$ Dept of Material Science and Engineering, University of Wisconsin, Madison, WI 53706}

\begin{abstract}
The inter-band optical absorption in graphene characterized by its fine-structure constant has a universal value of 2.3\% independent of the material parameters. However, for several graphene-based photonic applications, enhanced optical absorption in graphene is highly desired. In this work, we quantify the tunability of optical absorption in graphene via the Fermi level in graphene, angle of incidence of the incident polarized light, and the dielectric constant of the surrounding dielectric media in which graphene is embedded. The influence of impurities adsorbed on the surface of graphene on the Lorentzian broadening of the spectral function of the density of states is analytically evaluated within the equilibrium Green's function formalism. Finally, we compute the differential absorption of right and left circularly-polarized light in graphene that is uniaxially and optically strained. The preferential absorption or circular dichroism is investigated for armchair and zigzag strain. 
\end{abstract}
\maketitle
\section{Introduction}
The two-dimensional material graphene~\cite{neto2009electronic,allen2009honeycomb,geim2007rise}, which is a layer of carbon atoms arranged in a honeycomb lattice, exhibits strong light-matter interaction~\cite{koppens2011graphene,ju2011graphene} over a very wide wavelength ranging from the far infrared to the ultraviolet. The tunability of the density of states and the Fermi level in graphene along with its excellent transport properties reflected in a high carrier mobility~\cite{bolotin2008ultrahigh} provide a path for photonic applications such as quantum optics~\cite{pan2010hydrothermal,zhang2008tuning}, photo-voltaics~\cite{murray2011graphene,won2010photovoltaics}, photo-detectors~\cite{mueller2010graphene}, and biological sensing.~\cite{zhu2011strongly} 
While the optical characteristics of mono-layer graphene, attributed to enhanced light matter absorption with a high quantum efficiency, make it a desirable material for optical resonators and thermal-imaging cameras, the optical absorption is poor to be an efficient photo-detector.~\cite{kim2011role,liu2014graphene} Further, the adaptability of the optical absorption in graphene for various frequencies is limited by a flat absorption spectrum in the visible to the near-infrared region.~\cite{wang2008gate}  

In this work, we propose methods to improve the overall absorption in mono-layer graphene in a dielectric environment via a direct tuning of its optical conductivity. We begin with a description of the optical characteristics of a graphene sheet suspended in vacuum and derive the quantum of light absorption for a beam at oblique incidence (Section II A). The calculations are repeated for a graphene sheet sandwiched between two dielectrics; the dielectric function of graphene in each case is established using a standard RPA~\cite{bruus2004many} calculation. While most calculations tacitly assume an idealized set-up in which graphene is pristine thus preserving its electronic structure, especially the linearly dispersing bands around the $ K $ and $ K^{'}$ edges of the Brillouin zone, impurity atoms are usually adsorbed on the surface~\cite{zhou2007substrate,chen2009defect} to alter the electronic spectrum and attendant optical response. The correction  in this situation is manifest through a spectral broadening of the density of states and a changed absorption profile. The broadening for an impurity-adsorbed graphene sheet is evaluated by employing a Hubbard-type Hamiltonian within the equilibrium Green's function formalism (Section II B). Finally, in Section II C, a uniaxially- and optically-strained graphene sheet is analyzed for circular dichroism. Results are collected in Section III, and the paper concludes by briefly touching upon the implications of increased absorption for graphene-based optical devices.
\vspace{0.35cm}
\section{Optical absorption in suspended graphene}
\vspace{0.35cm}
Dispersion relationships for graphene bands with linearly dispersing eigen states~\cite{katsnelson2012graphene} are modeled using a two-dimensional Dirac Hamiltonian given as
\begin{equation}
\mathcal{H} = \hbar v_{f}(\sigma_{x}k_{y} - \sigma_{y}k_{x}) + \Delta\sigma_{z},
\label{dss}
\end{equation}
where $ v_{f}$ denotes the Fermi velocity of carriers, $\overrightarrow{k}$ is the wave-vector  measured relative to the Dirac points, and $\sigma_{i} $ ($ {i = x,y,z} $) are the usual Pauli matrices. The corresponding wave functions, in momentum space, for the momentum around the Dirac point have the form:
\begin{subequations}
\begin{equation}
\Psi_{\eta} = \dfrac{1}{\sqrt{2}}\begin{pmatrix}
u_{\eta}(k)\exp(-i\theta) \\
\eta u_{-\eta}(k)
\end{pmatrix},
\label{wfun1}
\end{equation}
where  $ \eta = \pm $, and $ u_{\eta} $ is given by
\begin{equation}
u_{\eta}(k) = \sqrt{1 \pm \dfrac{\Delta}{\sqrt{\Delta^{2}+ \left( \hbar v_{f}k\right)^{2}}}}.
\label{wfun2}
\end{equation}
\label{wfun_comp}
Here $ \Delta $ is the band gap induced in graphene, $ \theta_{k} = \arctan\left(\dfrac{k_{y}}{k_{x }}\right) $, and $ \eta = \pm $ corresponds to conduction and valence bands. The band gap introduced is primarily through an interaction with the substrate on which graphene is eptaxially grown, for instance, the honeycomb lattice rigidly held on the hexagonal BN~\cite{zhou2007substrate,dean2010boron} exhibits a band gap between 7.0 $ \mathrm{meV} $ to 20.0 $ \mathrm{meV}$.~\cite{jung2014origin} The band gap of pristine graphene, which hosts massless Dirac fermions, is zero.
\end{subequations}

\vspace{0.3cm}
\subsection{Transfer matrix for optical absorption}
\begin{figure}[t]
\includegraphics{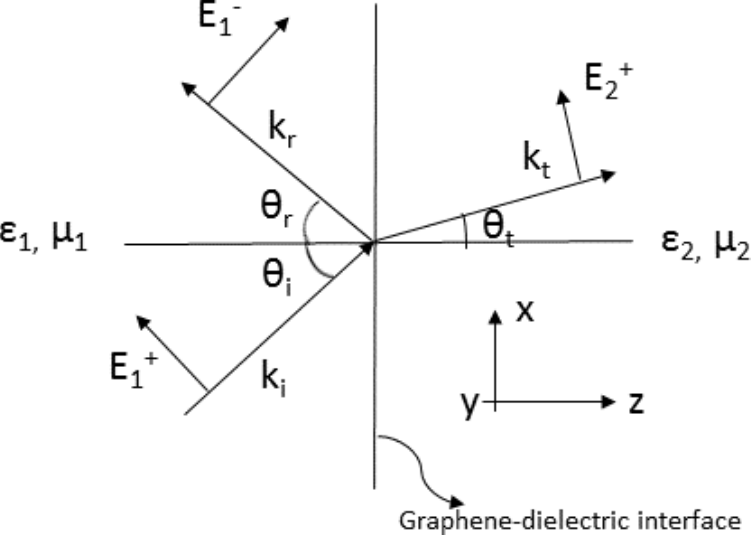} 
\caption{Schematic of a TM (\textit{p}-polarized) wave incident at angle $ \theta_{i}$ on the graphene-dielectric interface. The incident ray is partly reflected and partly transmitted. The angles of reflection and transmittance are indicated on the sketch. The electric field is normal to the propagation vector, while the magnetic field is along the \textit{y}-axis. It is assumed that there is no beam incident on the interface from medium 2.} 
\label{figs}
\end{figure}

\vspace{0.3cm}
To determine the optical response of carriers in graphene, linearly-polarized light along $x$-axis is assumed to shine perpendicularly on the graphene surface as indicated in Fig. \ref{figs}. 
The electric field of the light beam can be determined using the vector potential $ \overrightarrow{A}(t) = \overrightarrow{A}\exp(-i\omega t) $, where $\omega$ is the frequency of the incident light. The corresponding electric field is given as
\begin{equation}
\overrightarrow{E}(t) = -\dfrac{1}{c}\dfrac{\partial \overrightarrow{A}(t)}{\partial t}.
\label{efield}
\end{equation}
The speed of light is denoted by $ c $ in Eq.~\ref{efield}.
The Hamiltonian of Eq.~\ref{dss} can therefore be modified using the Peierls substitution~\cite{graf1995electromagnetic} and takes the form $ \mathcal{H} = \hbar v_{f}\overrightarrow{\sigma}.\left(\overrightarrow{k} - \dfrac{e}{c}\overrightarrow{A}(t)\right) $. We extract the interaction part of the Hamiltonian from the modified Hamiltonian and is expressed as
\begin{equation}
\mathcal{H}_{int} = -\dfrac{\hbar v_{f}e}{2c}\overrightarrow{\sigma}.\overrightarrow{A}(t).
\label{hint1} 
\end{equation} 
Substituting for $ \overrightarrow{A}(t) $ from Eq.~\ref{efield}, $ \mathcal{H}_{int} = \dfrac{i \hbar e v_{f}}{2\omega}\overrightarrow{\sigma}.\overrightarrow{E} $.
The factor of 0.5 in Eq.~\ref{hint1} comes from by retaining only the real part in the expansion of the magnetic vector potential $ \overrightarrow{A} $.

\noindent Using Fermi golden rule, the transition probability for light-induced transition from valence to conduction band is 
\begin{equation}
\dfrac{1}{\tau} = \sum\limits_{k_{c},k_{v}}\dfrac{2 \pi}{\hbar}\vert\langle \Psi_{f}\vert \mathcal{H}_{int}\vert\Psi_{i}\rangle\vert^{2}\delta\left(E_{c} - E_{v} - \hbar\omega\right). 
\label{fermis}
\end{equation}
Here, $\psi_f$ and $\psi_i$ correspond to the final and the initial scattering states, respectively; $E_c$ and $E_v$ refer to the energies at the bottom of the conduction band and the top of the valence band, respectively.

\noindent Recognizing the delta function in Eq.~\ref{fermis} as the density of states given as $ \dfrac{2\vert \varepsilon \vert}{\pi \hbar^{2}v_{f}^{2}} $ for the linear energy-dispersion relationship in graphene and integrating over energy space gives the absorbed flux (see Appendix A) as $ \dfrac{e^{2}\vert E \vert^{2}}{4\hbar} $. The incident flux~\cite{jackson1962classical} for an electric field is $ \dfrac{c}{4 \pi}\vert E\vert^{2} $. The optical inter-band absorption in graphene is, therefore, universally given as $ \dfrac{\pi e^{2}}{\hbar c} = 2.3 \% $, a material-independent number.

The optical absorption changes when graphene is sandwiched between two dielectric layers characterized by dielectric constants of $ \epsilon_{1} $ and $ \epsilon_{2}$. A transverse-magnetic~\cite{ramo2007fields} (TM or $p$ - polarized wave) with a magnetic field along the \textit{y}-axis (Eq.~\ref{bfield}) is assumed to impinge on the graphene-dielectric interface. The magnetic field corresponding to the TM polarization and propagating along the \textit{x}-axis is given as
\begin{equation}
{H}_{y,i} = \left(A_{i}e^{-ik_{z,i}z} + B_{i}e^{ik_{z,i}z}\right)e^{ik_{x}x},
\label{bfield}
\end{equation} 
where $ k_{z} = k_{in}\cos\theta $ and $ k_{x} = k_{in}\sin\theta $. A particular physical quantity within the \textit{i$^{th}$}layer in the structure is identified by the subscript \textit{i}, $ k_{in} $ is incident wave vector, and the angle of incidence is $ \theta $. The appropriate boundary conditions at the interface \textit{z} = 0 are
\begin{subequations}
\begin{equation}
\widehat{n}\times \left(\overrightarrow{E}_{j} - \overrightarrow{E}_{i}\right) = 0,
\label{ebc}
\end{equation}
and
\begin{equation}
\widehat{n} \times \left(\overrightarrow{H}_{j} - \overrightarrow{H}_{i}\right) = \overrightarrow{J}_{surf},
\label{hbc}
\end{equation}
where $ \overrightarrow{J}_{surf}$ is the surface current due to the 2D charge carriers in the graphene sheet. The subscripts \textit{i}, \textit{j} denote a particular material layer on the left and the right side of a given interface, respectively. The electric field can be obtained using Maxwell's equation $ \overrightarrow{\nabla} \times \overrightarrow{E} = -\dfrac{\partial \overrightarrow{B}}{\partial t} $ to yield $ \overrightarrow{E}_{i} = -\dfrac{i}{\omega \varepsilon_{0}\varepsilon_{i}} \overrightarrow{\nabla} \times \overrightarrow{H}$.
\end{subequations}
For the form of the magnetic field chosen in Eq.~\ref{bfield}, the corresponding electric field is given as
\begin{subequations}
\begin{equation}
E_{z,i} = \frac{k_x}{\omega\epsilon_0\epsilon_i}\left(A_ie^{-ik_{z,i}z}+B_ie^{ik_{z,i}z}\right)e^{ik_xx},
\end{equation}
\begin{equation}
E_{x,i} = \frac{k_{z,i}}{\omega\epsilon_0\epsilon_i}\left(A_ie^{-ik_{z,i}z}-B_ie^{ik_{z,i}z}\right)e^{ik_xx}.
\end{equation}
\label{efield1}
\end{subequations}
Combining the above equations with the boundary condition (Eq.~\ref{ebc}) for electric field on either side of the graphene-dielectric interface gives
\begin{subequations}
\begin{equation}
\dfrac{k_{z,1}}{\epsilon_{1}}\left(A_{1} - B_{1}\right) - \dfrac{k_{z,2}}{\epsilon_{2}}\left(A_{2} - B_{2}\right) = 0.
\label{ebc1}
\end{equation}
The second relation using boundary conditions applied on the magnetic field (Eq.~\ref{hbc}) is
\begin{eqnarray}
J_x &=& \left(A_{1} + B_{1}\right) - \left(A_{2} + B_{2}\right), \\
&=& \sigma\dfrac{k_{z,2}\left(A_{2}-B_{2}\right)}{\epsilon_{0}\epsilon_{2}}.
\label{bbc2}
\end{eqnarray}
\end{subequations}
Note that in the above equation, $\sigma$ is the complex dynamical conductivity of graphene as discussed in Section II B. 
All quantities on left(right) of the interface are subscripted as ``1(2)''. In matrix notation, the amplitude of the magnetic fields on either side of the interface are related as~\cite{markos2008wave}
\begin{subequations}
\begin{equation}
\begin{pmatrix} A_{2} \\
B_{2}
\end{pmatrix} =  M_{tm} \begin{pmatrix} A_{1} \\
B_{1}
\end{pmatrix}.
\label{tmat1}
\end{equation}
The transfer matrix $ M_{tm} = M_{1,tm} + M_{2,tm} $, where
\begin{equation}
M_{1,tm} = \frac{1}{2}\begin{pmatrix}
1 + \dfrac{\epsilon_{1}k_{z,2}}{\epsilon_{2}k_{z,1}} & 1 - \dfrac{\epsilon_{1}k_{z,2}}{\epsilon_{2}k_{z,1}} \\
1 - \dfrac{\epsilon_{1}k_{z,2}}{\epsilon_{2}k_{z,1}} & 1 + \dfrac{\epsilon_{1}k_{z,2}}{\epsilon_{2}k_{z,1}}
\end{pmatrix},
\label{m1tm}
\end{equation}
and 
\begin{equation}
M_{2,tm} = \frac{1}{2}\begin{pmatrix}
\dfrac{\sigma k_{z,2}}{\epsilon_{0}\epsilon_{2}\omega} & -\dfrac{\sigma k_{z,2}}{\epsilon_{0}\epsilon_{2}\omega} \\
\dfrac{\sigma k_{z,2}}{\epsilon_{0}\epsilon_{2}\omega} & -\dfrac{\sigma k_{z,2}}{\epsilon_{0}\epsilon_{2}\omega}
\end{pmatrix}.
\label{m2tm}
\end{equation}
\end{subequations}

\noindent The transfer matrix $ M_{tm} $ derived in Eq.~\ref{tmat1} allows us to compute reflectance and transmittance amplitudes; in particular, reflectance and transmittance amplitudes are
\begin{eqnarray}
r = \dfrac{M_{tm}\left(2,1\right)}{M_{tm}\left(1,1\right)}, \label{ref} \\
t = \dfrac{1}{M_{tm}\left(1,1\right)}.
\label{trans}
\end{eqnarray}
The reflection and transmission coefficients are $ R = r^{2} $ and $ T = t^{2} $ which add to unity $\left(R + T = 1 \right)$  for zero absorption losses.

At this point it is imperative to discuss appropriate limits that are placed on the expressions for reflectance and transmittance amplitudes derived in Eqs.~\ref{ref} and~\ref{trans}. The amplitudes must satisfy $ 0 \leq r \leq 1 $ and $ 0 \leq t \leq 1 $ from which follows 
\begin{subequations}
\begin{equation}
0 \leq t = \dfrac{1}{M_{tm}\left(1,1\right)} \leq 1.
\label{trel1}
\end{equation}
To evaluate the constraint, we note that the matrix $ M_{tm,2} $ which describes the optical conductivity of the graphene sheet makes a small contribution to the overall matrix $ M_{tm} $ in the high-frequency limit. We are, therefore, left with the inequality
\begin{equation}
0 \leq t = \dfrac{2\epsilon_{2}k_{1z}}{\epsilon_{2}k_{1z} + \epsilon_{1}k_{2z}} \leq 1.
\label{trel2}
\end{equation}
Since all quantities in the above equation are assumed to be positive, the left half side of the above inequality is trivially true; evaluating the right hand side inequality, one arrives at the relation $ \dfrac{\epsilon_{2}}{\epsilon_{1}} \leq \dfrac{k_{2}}{k_{1}} $. For a case which is in violation of this condition, unphysical solutions are obtained. We also demonstrate a parallel condition by placing a similar constraint on the reflectance amplitude. Proceeding as above, we write
\begin{equation}
0 \leq  r = \dfrac{M_{tm}\left(2,1\right)}{M_{tm}\left(1,1\right)} \leq 1
\label{rel1}
\end{equation}
Ignoring the contribution of $ M_{tm,2} $ in the high-frequency limit, the right inequality $ \left( Eq.~\ref{rel1} \right) $ is trivially satisfied while the left takes the form
\begin{equation}
r = \dfrac{\epsilon_{2}k_{1z}- \epsilon_{1}k_{2z}}{\epsilon_{2}k_{1z} + \epsilon_{1}k_{2z}}.
\label{rel2}
\end{equation}
It is straightforward to see that the condition $ r \geq 0 $ leads us to $ \dfrac{\epsilon_{2}}{\epsilon_{1}} > \dfrac{k_{2}}{k_{1}} $, which means the transmittance amplitude is greater than unity. This apparent contradiction is resolved by noting that in this limiting case $ \dfrac{\epsilon_{2}}{\epsilon_{1}} = \dfrac{k_{2}}{k_{1}} $, the reflected component is zero and the incident beam is fully transmitted $ \left( t = 1\right) $. We have thus arrived at an analogous condition for obtaining the corresponding Brewster' angle for a graphene sheet embedded in inhomogeneous dielectric media. Of course, the limits on the validity of the ratio $\epsilon_2/\epsilon_1$ must also account for the finite contribution from the transfer matrix $M_{2,tm}$ in the lower frequency regime. 
\end{subequations} 

\noindent The transfer matrix $ M_{te} $ for a TE~\cite{ramo2007fields} wave or $ s $ - polarized wave can be similarly derived using the appropriate Maxwell's boundary conditions. The matrix $ M_{te} = M_{1,te} + M_{2,te} $ is
\begin{subequations}
\begin{equation}
M_{1,te} = \dfrac{1}{2}\begin{pmatrix}
1 + \dfrac{\mu_{1}k_{z,2}}{\mu_{2}k_{z,1}} & 1 - \dfrac{\mu_{1}k_{z,2}}{\mu_{2}k_{z,1}} \\
1 - \dfrac{\mu_{1}k_{z,2}}{\mu_{2}k_{z,1}} & 1 + \dfrac{\mu_{1}k_{z,2}}{\mu_{2}k_{z,1}}
\end{pmatrix},
\label{m1te}
\end{equation}
and 
\begin{equation}
M_{2,te} = \dfrac{1}{2}\begin{pmatrix}
\dfrac{\mu_{0}\mu_{1}\omega\sigma}{k_{1,z}} & \dfrac{\mu_{0}\mu_{1}\omega\sigma}{k_{1,z}} \\
-\dfrac{\mu_{0}\mu_{1}\omega\sigma}{k_{1,z}} & -\dfrac{\mu_{0}\mu_{1}\omega\sigma}{k_{1,z}}
\end{pmatrix}.
\label{m2te}
\end{equation}
\end{subequations}
The magnetic permeabilities $\left( \mu_{1},\mu_{2}\right) $ are taken to be unity for non-magnetic dielectric media.

\vspace{0.35cm}
\subsection{Optical absorption in impure graphene}
\vspace{0.35cm}
Electromagnetic absorption in graphene is determined by its complex dynamical conductivity.
The dynamical conductivity of graphene is obtained from the dielectric function $ \epsilon\left(q,\omega\right)$ through a random phase approximation (RPA). In the long-wavelength limit $\left(q \rightarrow 0\right)$, the RPA dielectric function is given as~\cite{hwang2007dielectric,hill2009dielectric,bludov2013primer}  
\begin{align}
\begin{split}
\epsilon\left( q \rightarrow 0, \omega\right) = 1 - \dfrac{2 \pi e^{2}}{q}\dfrac{q^{2}}{2 \pi \hbar \omega}\left[\dfrac{2E_{f}}{\hbar \omega} + \dfrac{1}{2}\ln\left\vert \dfrac{ 2E_{f}- \hbar \omega}{2E_{f} + \hbar \omega}\right\vert \right. \\
\left. - i\dfrac{\pi}{2}\Theta\left(\hbar \omega - 2E_{f} \right)\right], 
\label{dcgp}
\end{split} 
\end{align}
where $E_f$ is the Fermi level in the graphene sheet, and $\Theta(.)$ is the Heaviside function. The dynamical conductivity, $\sigma(q,\omega)$, is related to the dielectric constant as
\begin{equation}
\sigma\left(q,\omega\right) = \dfrac{i \omega}{2 \pi q}\left[1 - \epsilon\left(q,\omega\right)\right]. 
\label{coneps} 
\end{equation}
Inserting Eq.~\ref{dcgp} in Eq.~\ref{coneps} gives
\begin{equation}
\begin{split}
\sigma\left(q \rightarrow 0, \omega\right) = \dfrac{E_{f}e^{2}}{\pi \hbar}\dfrac{i}{\hbar \omega + i\Gamma} + \dfrac{e^{2}}{4\hbar}\Theta\left(\hbar \omega - 2E_{F}\right) \\
+ \dfrac{ie^{2}}{4 \pi \hbar}\ln\left\vert \dfrac{ 2E_{f}- \hbar \omega}{2E_{f} + \hbar \omega}\right\vert. 
\label{comcon} 
\end{split}
\end{equation}
The spectral width $ \Gamma $ corresponds to a Lorentzian broadening of the density of states~\cite{duke1965phonon,christen1990line} and is determined by the impurities adsorbed on the graphene surface. In this work, using a retarded Green's function approach, we evaluate the spectral width, $\Gamma$. A model Hubbard Hamiltonian~\cite{jishi2013feynman} that captures an atom adsorbed on graphene can be written as
\begin{flalign}
\begin{split}
\mathcal{H} =\sum\limits_{k}\varepsilon_{gr}a^{\dagger}_{gr}a_{gr} + \sum\limits_{s}\varepsilon_{ad}c^{\dagger}_{s}c_{s} + \sum\limits_{k}V_{hyb}a^{\dagger}_{gr}c_{s}  \\
+ \sum\limits_{k}V_{hyb}^{*}c^{\dagger}_{s}a_{gr}. 
\label{hamhubb}
\end{split}
\end{flalign}
This Hamiltonian resembles the non-interacting Anderson impurity model for resonant impurities.~\cite{wehling2009adsorbates}
The energy of the adsorbed atom is given by $ \varepsilon_{ad} $, while $ \varepsilon_{gr} $ denotes the energy of the graphene electron states. The last two terms of the Hamiltonian describe the hybridization between the adsorbed atom and graphene. The summation over momentum vectors also denote the dual spin states. The creation (annihilation) operators for graphene and the adsorbed atom are denoted $ a^{\dagger} (a) $ and $ c^{\dagger}_{s} (c_{s}) $, respectively. 

To evaluate the spectral density function from which the overall broadening is determined, we consider the graphene-adsorbed impurity to be a non-interacting system. Using the equation of motion approach within the equilibrium Green's function formalism, we start by writing the retarded Green's function~\cite{mahan2000many} for the adsorbed atom 
\begin{equation}
G^{R}\left(c,s,t\right) = -i\theta\left(t\right) \langle\lbrace c_{s}\left(t\right), c_{s}^{\dagger}\left(t\right)\rbrace\rangle.
\label{gr1}
\end{equation}
Taking the derivative of the retarded Green's function in Eq.~\ref{gr1} and following Heisenberg's picture $\left(\mathcal{H}\left(t\right)= \exp\left(i\mathcal{H}t\right)\mathcal{H}\exp\left(-i\mathcal{H}t\right)\right) $ gives
\begin{equation}
i\dfrac{\partial}{\partial t}G^{R}\left(c,s,t\right) = \delta\left(t\right) + \dfrac{i}{\hbar}\theta\left(t\right)\langle\lbrace\left[ \mathcal{H}, c_{s}\left(t\right)\right] ,  c_{s}^{\dagger}\left(t\right)\rbrace\rangle.
\label{gr2}
\end{equation}
Expanding Eq.~\ref{gr2} by evaluating the commutator (see Appendix B), one obtains
\begin{equation}
i\dfrac{\partial}{\partial t}G^{R}\left(c,s,t\right) = \delta\left(t\right) + \dfrac{\varepsilon_{d}}{\hbar}G^{R}\left(c,s,t\right) + \dfrac{1}{\hbar}\sum\limits_{k}V_{hyb}G^{R}\left(k,t\right),
\label{gr3}
\end{equation}
where $ G^{R}\left(k,t\right) $ is the retarded Green's function for graphene-adsorbed atom. $ G^{R}\left(k,t\right) $ can be written in standard form as
\begin{equation}
G^{R}\left(k,t\right) = - i\theta\left(t\right)\langle\lbrace a_{gr}\left(t\right), c_{s}^{\dagger}\left(0\right)\rbrace\rangle.
\label{gr4}
\end{equation}
The corresponding equation of motion is
\begin{equation}
i\dfrac{\partial}{\partial t}G^{R}\left(k,t\right) = \dfrac{\varepsilon_{gr}}{\hbar}G^{R}\left(k,t\right) + \dfrac{1}{\hbar}V_{hyb}G^{R}\left(c,s,t\right).
\label{gr5}
\end{equation}
The Fourier transform (Eq.~\ref{gr6}) of the two retarded Green's functions yield a pair equations (Eq.~\ref{gr7},~\ref{gr8}) which can be solved for $ G^{R}\left(c, \omega\right) $ 
\begin{equation} 
G^{R}\left(c,s,t\right) = \dfrac{1}{2 \pi}\int_{-\infty}^{\infty}e^{-i\omega t}G^{R}\left(c,s,\omega\right),
\label{gr6}
\end{equation}
\begin{equation}
\left(\omega - \dfrac{\varepsilon_{ad}}{\hbar} G^{R}\left(c,\omega\right)\right) = 1 + \dfrac{1}{\hbar}V_{hyb}^{*}G^{R}\left(k,\omega\right),
\label{gr7}
\end{equation}
and
\begin{equation}
\left(\omega - \dfrac{\varepsilon_{gr}}{\hbar} G^{R}\left(k,\omega\right)\right) = \dfrac{1}{\hbar}V_{hyb}G^{R}\left(c,s,\omega\right).
\label{gr8}
\end{equation}
Solving for $ G^{R}\left(c,\omega\right) $ and preserving causality by making the substitution $ \omega \rightarrow \omega + i0^{+} $ gives
\begin{flalign}
G^{R}\left(c,s,\omega\right) = \dfrac{\hbar}{\hbar \omega + i0^{+} - \varepsilon_{ad} - \sum\limits_{k}\dfrac{\vert V_{hyb}\vert^{2}}{\hbar \omega - \varepsilon_{gr}+ i0^{+}}} \notag \\
= \dfrac{\hbar}{\hbar \omega - \varepsilon_{ad} - \sum\limits_{k}P\dfrac{\vert V_{hyb}\vert^{2}}{\hbar \omega - \varepsilon_{gr}}+ i\pi\sum\limits_{k}\vert V_{hyb}\vert^{2}\delta\left(\hbar \omega - \varepsilon_{gr}\right)} \notag \\
\end{flalign}
where $ P $ stands for the principal value~\cite{hassani2008mathematical} in the usual Plemelj relation: $ \dfrac{1}{x \pm i\varepsilon} = P\left(\dfrac{1}{x}\right) \pm i\pi\delta\left(x\right) $. The delta function is again the usual density of states for graphene; putting all of them together and evaluating the spectral density $ A\left(c, s,\omega\right) = -2Im{G^{R}\left(c,s,\omega\right)} $ gives
\begin{flalign}
A\left(c,s,\omega\right) =  \dfrac{2 \hbar \Gamma}{\left[\hbar \omega - \varepsilon_{ad}-\sum\limits_{k}P\dfrac{\vert V_{hyb}\vert^{2}}{\hbar \omega - \varepsilon_{gr}}\right]^2 + \Gamma^{2}},
\label{gr9}
\end{flalign}
where $ d\left(E_{f}\right) = \dfrac{2\vert E_f \vert}{\pi \hbar^{2}v_{f}^{2}} $ is the density of states for graphene at the Fermi level and the broadening parameter $ \Gamma = \pi S \vert V_{hyb} \vert^{2}d\left(E_f\right)$; $S$ denotes the area of the graphene sheet under consideration and is roughly given as $S \approx 1/n_{imp}$, where $n_{imp}$ is the concentration of impurity atoms adsorbed on the graphene sheet. The hybridization potential can be estimated as $V_{hyb} \approx \hbar v_f/\left(R_1\sqrt{|\ln(R_0/R_1)|}\right)$.~\cite{rakheja2013evaluation} Here, $R_0$ is the radius of the impurity atom, and $R_1$ is the average distance between the adsorbed impurity atoms and is approximately given as $1/\sqrt{n_{imp}}$.
 
\noindent Impurities therefore broaden the Dirac-delta peak, the broadening given by the strength of the hybridization potential, $ V_{hyb} $. We have tacitly assumed that the system is non-interacting, and  there is no Coulomb repulsion term of the form $ U n_{d,\uparrow} n_{d,\downarrow} $ that appears in a standard Hubbard Hamiltonian. 
\vspace{0.35cm}

\subsection{Circular dichroism in strained graphene}
\vspace{0.35cm}
Circular dichroism quantifies the differential absorption of right and left circularly-polarized light. The expression for absorption coefficient due to light-induced inter-band transitions is a measure of the strength of the optical matrix element.~\cite{vasko1999electronic,patterson2007solid} Inter-band optical absorption from a populated valence band eigen state to an empty conduction band state requires the determination of the optical inter-band transition matrix elements $ \vert P^{+}_{cv}\vert^{2} $ and $ \vert P^{-}_{cv}\vert^{2} $ for right and left circularly-polarized light,respectively. $ P_{cv}^{x} $, for instance, for right circularly polarized light is nominally defined as~\cite{li2014hexagonal}
\begin{equation}
P_{cv}^{x} = \langle\Psi_{+}\vert \triangledown_{k_{x}}\mathcal{H}\vert\Psi_{-}\rangle.
\label{optmatr}
\end{equation} 
The degree of circular polarization~\cite{berova2000circular} $ \rho\left(k\right)$ is therefore
\begin{equation}
\rho\left(k\right) = \dfrac{\vert P^{+}_{cv}\vert^{2} - \vert P^{-}_{cv}\vert^{2}}{\vert P^{+}_{cv}\vert^{2} + \vert P^{-}_{cv}\vert^{2}}. 
\label{dichro}
\end{equation}
The inter-band matrix elements for right and left circularly-polarized light, $ \vert P^{+}_{cv}\vert $ and $ \vert P^{-}_{cv}\vert $, respectively are defined as
\begin{equation}
 P^{\pm}_{cv} =  P_{cv}^{x} \pm iP_{cv}^{y}
\label{defdc}
\end{equation} 

We utilize Eq.~\ref{optmatr} to compute the inter-band optical matrix elements in strained graphene and establish a relation between dischroism and the tunable Fermi level. A graphene sheet usually experiences acoustic and optical strain. Under acoustic strain, the two carbon atoms of the unit cell are displaced together, while for the optical case, there is a shift such that the centre-of-mass remains invariant. An out-of-plane component of optical strain renders the two sub-lattices of graphene inequivalent, a situation commonly realized when graphene is grown on a substrate, for instance, boron nitride. The Hamiltonian using the method of invariants that describes this situation is written as~\cite{linnik2012effective}
\begin{flalign}
\mathcal{H}_{str} &= \hbar v_{f}\left\lbrace \left(k_{x} - ik_{y}\right)\sigma_{+} + \left(k_{x} + ik_{y}\right)\sigma_{-}\right\rbrace + E_{op}\sigma_{z}, \notag \\
&= \hbar v_{f}\begin{pmatrix}
E_{op} & k_{x} - ik_{y} \\
k_{x} + ik_{y} & -E_{op}
\end{pmatrix},
\label{hamgrs} 
\end{flalign}
where $ \sigma_{\pm} = \left(\sigma_{x} \pm i\sigma_{y}\right)/2 $ and $ E_{op} $ is the optical strain. A gap, $ \Delta = 2E_{op} $, in the spectrum appears now which implies the non-equivalence of the graphene sub-lattices. Interestingly, the out-of-plane optical strain component, while it breaks the reflection symmetry, does not move the Dirac points from the Brillouin zone edges $ K $ and $ K^{'} $.

The optical matrix elements and the tunable circular dichroism of an optically strained graphene sheet can be calculated by first working out the velocity components.
The \textit{x} and \textit{y} components of velocity are given by
\begin{subequations}
\begin{equation}
\dfrac{\partial \mathcal{H}_{str}}{\partial k_{x}} = \begin{pmatrix}
0 &  \hbar v_{f} \\
 \hbar v_{f} & 0
\end{pmatrix},
\end{equation}
and
\begin{equation}
\dfrac{\partial \mathcal{H}_{str}}{\partial k_{y}} = \begin{pmatrix}
0 & -i\hbar v_{f} \\
i\hbar v_{f} & 0
\end{pmatrix}.
\end{equation}
\label{velcomp}
\end{subequations}

\noindent The optical matrix element corresponding to the velocity components for a finite band gap graphene can therefore be written by inserting expressions for wave functions and the velocity components from Eq.~\ref{wfun_comp} and Eq.~\ref{velcomp}, respectively, in Eq.~\ref{optmatr} for a right circularly-polarized light. 
\begin{equation}
P_{cv,x} = \dfrac{1}{2}\begin{pmatrix}
u_{+}\exp\left(i\theta\right) & u_{-}  
\end{pmatrix} \begin{pmatrix}
0 & -i\hbar v_{f} \\
i\hbar v_{f} & 0
\end{pmatrix} \begin{pmatrix}
u_{-}\exp\left(-i\theta\right) \\
-u_{+}
\end{pmatrix}.
\label{pcvx}
\end{equation}
The matrix element corresponding to momentum operator $ \hat{p_{y}} $ is
\begin{equation}
P_{cv,y} = \dfrac{1}{2}\begin{pmatrix}
u_{+}\exp\left(i\theta\right) & u_{-}  
\end{pmatrix} \begin{pmatrix}
0 & \hbar v_{f} \\
\hbar v_{f} & 0
\end{pmatrix} \begin{pmatrix}
u_{-}\exp\left(-i\theta\right) \\
-u_{+}
\end{pmatrix}.
\label{pcvy}
\end{equation}
\label{pcvtot}
Combining both the components, the square of the right polarized optical matrix element is
\begin{equation}
\vert P^{+}_{cv}\vert^{2} = \left(\hbar v_{f}\right)^{2}\left[1 + \dfrac{\Delta}{\left(\hbar v_{f}k\right)^{2}+\Delta^{2}}\right]^{2}.
\label{pcvr2}  
\end{equation}
The corresponding expression for $ \vert P^{-}_{cv}\vert^{2}$ following an analogous procedure is
\begin{equation}
\vert P^{-}_{cv}\vert^{2} = \left(\hbar v_{f}\right)^{2}\left[1 + \dfrac{-\Delta}{\left(\hbar v_{f}k\right)^{2}+\Delta^{2}}\right]^{2}.
\label{pcvl2}  
\end{equation}

The above derivation for dichroism in graphene was carried out by considering an optical strain that furnishes a finite band gap. First principle calculations~\cite{choi2010effects} also show that an application of a large uniaxial strain in graphene does not destroy the semi-metallic nature of graphene but significantly impacts the Fermi velocity components. The anisotropy of Fermi velocity which manifests as tilted Dirac cones has been profitably employed to tune the optical properties of two-dimensional graphene.~\cite{pellegrino2010strain} We work out an expression for circular dichroism in graphene by considering uniaxial strain along the zigzag and armchair directions~\cite{brey2006electronic} in presence of a finite band gap. Writing out the Hamiltonian (Eq.~\ref{hamgrs}) again but with different velocity components, $ \mathcal{H} = v_{x}p_{x}\sigma_{x} + v_{y}p_{y}\sigma_{y} $, such that $ \kappa = v_{y}/v_{x}$. The ratio is usually computed by resorting to a single-orbital tight-binding calculation which expresses the velocity components as~\cite{sharma2012interacting}
\begin{flalign}
v_{x} &= t_{2}a_{x}\sqrt{4\eta^{2}-1} ,\notag \\
v_{y} &= t_{2}a_{y}.
\label{vanis}
\end{flalign}
Here, $ t_{i} \left(i = 1, 2,3\right) $ denotes the kinetic energy hopping integrals between the nearest neighbours, and $ \eta = t_{1}/t_{2} $. The strained lattice vector is resolved in $ x $ and $ y $ components to give and $ a_{x} = a/2 $ and $ a_{y} = \dfrac{\sqrt{3}a}{2} $, with $ a  = \sqrt{3} \times 1.42 \AA $. Using first principles calculations~\cite{choi2010effects}, it has been demonstrated that a uniaxial strain along the zig-zag direction (denoted as ``Z'' strain) in the honeycomb lattice leads to $t_1$ = $t_3$ $<$ $t_2$ ( $\eta < 1$), while for uniaxial strain in the armchair chain direction (denoted as ``A'' strain), $t_1$ = $t_3$ $>$ $t_2$ ($\eta > 1$). 

The degree of velocity anisotropy, $ \kappa $, using Eq.~\ref{vanis} is given as
\begin{equation}
\kappa = \dfrac{v_{y}}{v_{x}} = \sqrt{\dfrac{3}{\left(4\eta^{2}-1\right)}}. 
\label{vanis1}
\end{equation}

\noindent The degree of circular polarization (Eq.~\ref{dichro}) is modified to reflect this by an alteration to the algebraic expressions for both right and left polarized optical matrix elements. Carrying out the derivation as before gives
\begin{equation}
\vert P^{\pm}_{cv}\vert^{2} = \left(\hbar v_{r}\right)^{2}\left[1 + \dfrac{\pm \Delta}{\sqrt{\left(\hbar v_{x}k_{x}\right)^{2} + \left(\hbar v_{y}k_{y}\right)^{2}
+ \Delta^{2}}}\right]^{2},  
\end{equation}
where $ v_{r} = \sqrt{v_{x}^{2} + v_{y}^{2}} $. In terms of the velocity anisotropy, the above equation can be simplified as 
\begin{eqnarray}
\vert P^{\pm}_{cv}\vert^{2} = \left(\hbar v_{x}\right)^{2}\left(1+\kappa^2\right)\times \notag \\
\left[1 + \dfrac{\pm \Delta}{\sqrt{\left(\hbar v_{x}k_{x}\right)^{2}\left(1+\kappa^2\tan^2\theta_{k}\right)
+ \Delta^{2}}}\right]^{2},
\end{eqnarray}
where $\tan\theta_{k} = k_y/k_x $.

\noindent The average degree of circular polarization over a constant energy surface (free of trigonal warping effects~\cite{pereira2009valley,saito2000trigonal}) can be defined as
\begin{equation}
\rho\left(k\right) = \dfrac{\int d^{2}k \rho\left(k\right)\delta\left(\omega - 2\varepsilon\left(k\right)\right)}{\int d^{2}k \delta\left(\omega - 2\varepsilon\left(k\right)\right)}, 
\label{avgrho}
\end{equation}
where $ \varepsilon\left(k\right) =   \pm\sqrt{\hbar^{2}v_{r}^{2}k^{2}+ \Delta^{2}}$ is the energy spectrum of graphene. The frequency of the light beam must satisfy $ \omega = 2\varepsilon\left(k\right) $ which describes the energy involved in an inter-band transition from valence to conduction band.

\vspace{0.35cm}
\section{Results}
\begin{figure}[h]
\includegraphics[width=3.0in]{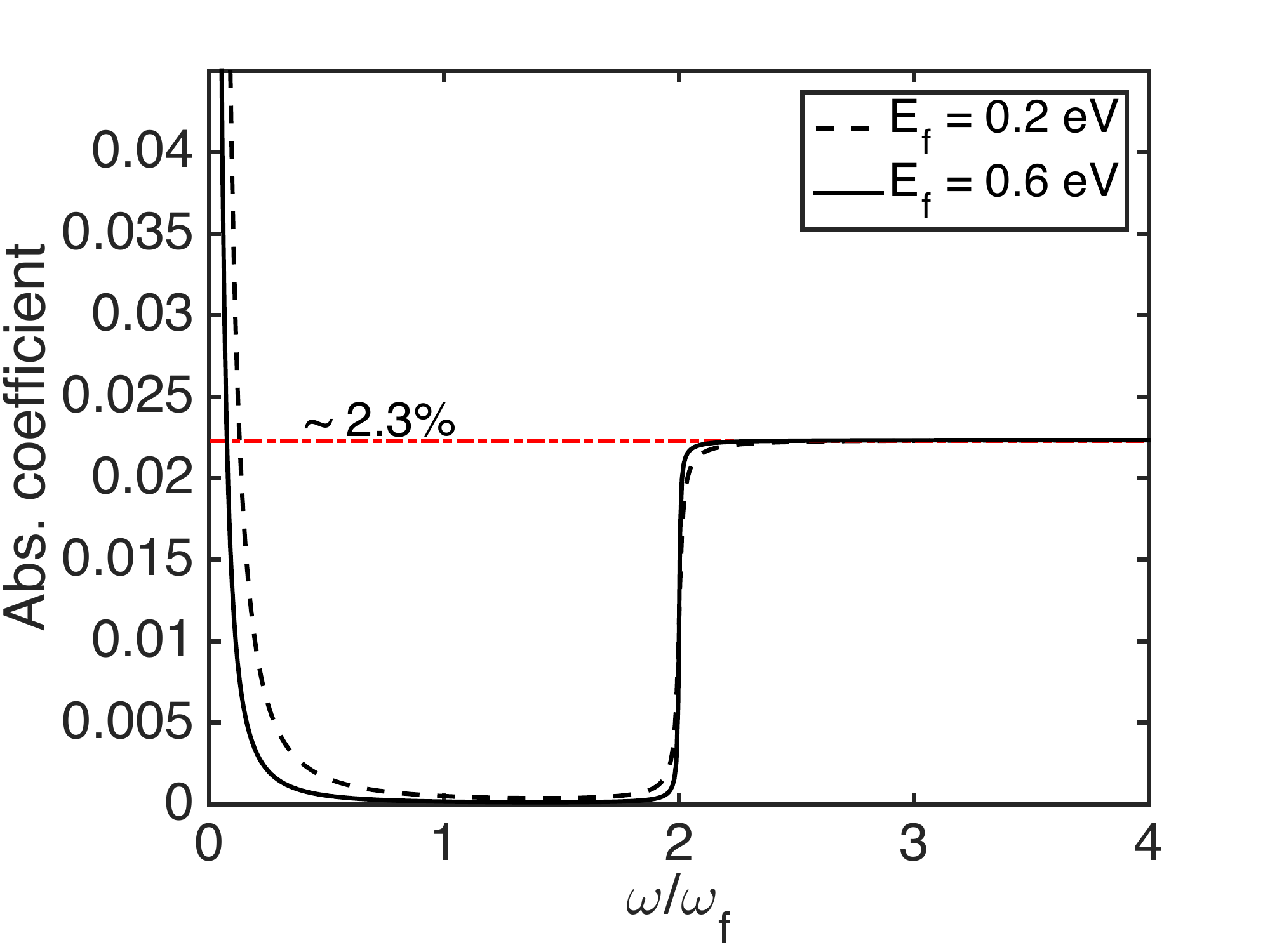} 
\caption{The absorption coefficient of mono-layer graphene suspended in vacuum is plotted for various energies. Here, $\omega_f$ is the frequency corresponding to the Fermi level in the graphene sheet. As shown, a constant 2.3\% absorption is obtained independent of the material parameters for $\omega > \omega_f$. While the Fermi level, $E_f$, plays a role in determining absorption in the low-frequency regime where $\omega << \omega_f$, in the high-frequency regime characterized by inter-band scatterings, absorption coefficient becomes independent of $E_f$.}
\label{fig1}
\end{figure}

\begin{figure}[h!]
\includegraphics[width=3.5in]{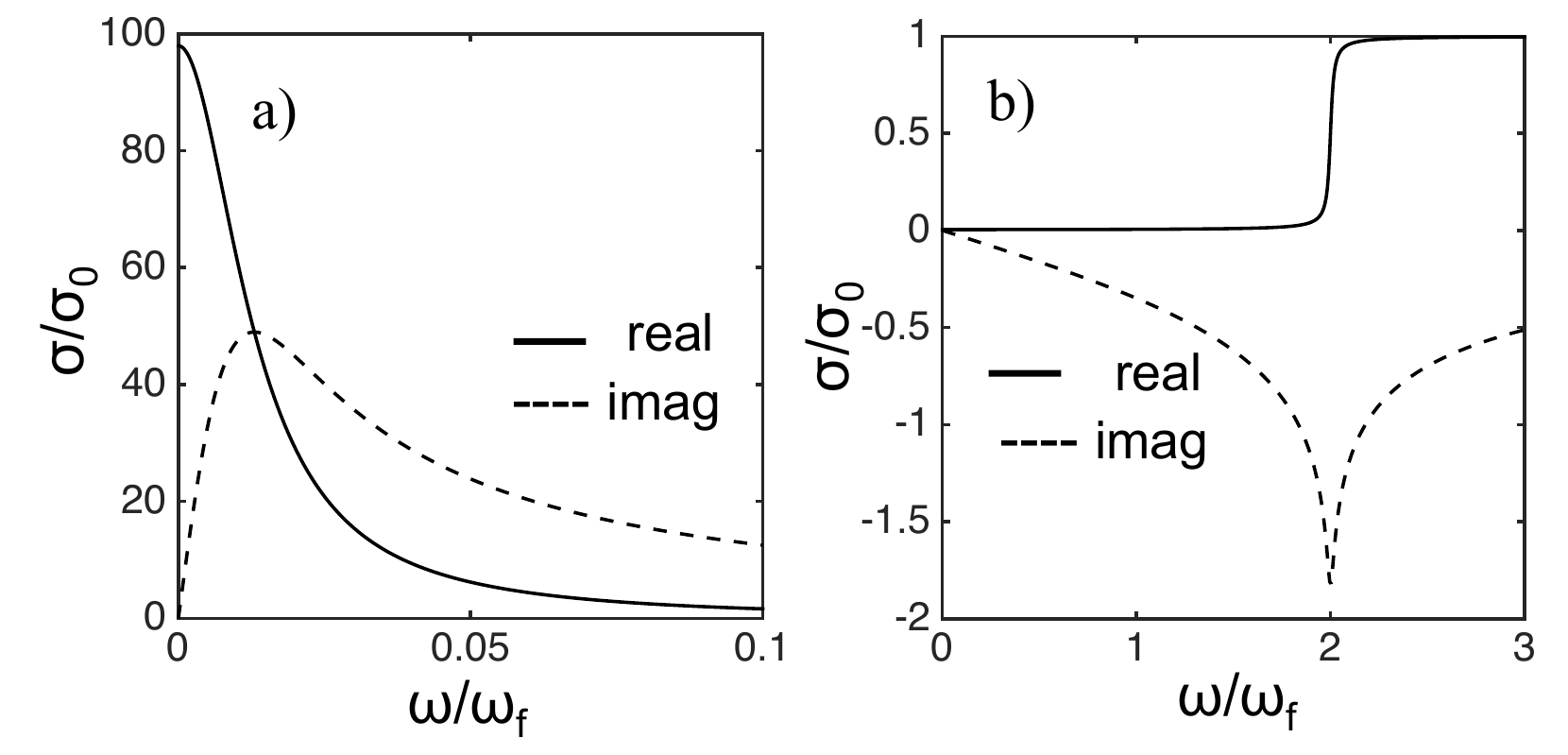} 
\caption{Complex dynamical conductivity of mono-layer graphene. The Fermi level is set to 0.2 $ \mathrm{eV}$. Fig.~\ref{fig1}a plots the Drude (intra-band) component of conductivity, while the inter-band component is shown in Fig.~\ref{fig1}b. The damping factor is assumed to be 2.6 $ \mathrm{meV} $. The real (imaginary) part of each contribution is a solid (dashed) line in each sub-figure. The $ x $-axis is normalized to frequency corresponding to the Fermi level. The $ y $-axis is normalized to $ \sigma_{0} = \pi e^{2}/2h $.} 
\label{fig2}
\end{figure}

\vspace{0.35cm}
As a first check of the validity of the numerical model to compute absorption, an undoped graphene sheet with an electrically tuned Fermi level and suspended in vacuum (the dielectric constants flanking the graphene sheet are unity) is considered. The absorption coefficient, at normal angle of incidence, for a range of energies is plotted in Fig.~\ref{fig1}. Two features in Fig.~\ref{fig1} stand out; when incident radiation energy is upwards of twice the Fermi level, the absorption is constant at approximately 2.3 $\%$ and the absorption coefficient has a ``hump'' at exactly $ \hbar \omega = 2E_{f} $. The numerical demonstration of absorption coefficient as approximately 2.3 $\%$ is significant since it is in conformity with the theoretically derived number (see Appendix A) obtained directly from the wave functions of graphene's linear Hamiltonian coupled to the incident electromagnetic field. The ``hump'' is attributed to an onset of inter-band conductivity at $ \hbar \omega = 2E_{f} $, beyond which the absorption coefficient stays constant. The intra-band Drude contribution is dominant at much lower energies as can be seen by observing the optical conductivity versus energy plot of Fig.~\ref{fig2}. The ``tuning'' of absorption coefficient via the optical conductivity of graphene will concern us for rest of this section.

\vspace{0.25cm}
\subsection{Absorption coefficient versus incident angle and material parameters}
\begin{figure} [h!]
\includegraphics[width=3.0in]{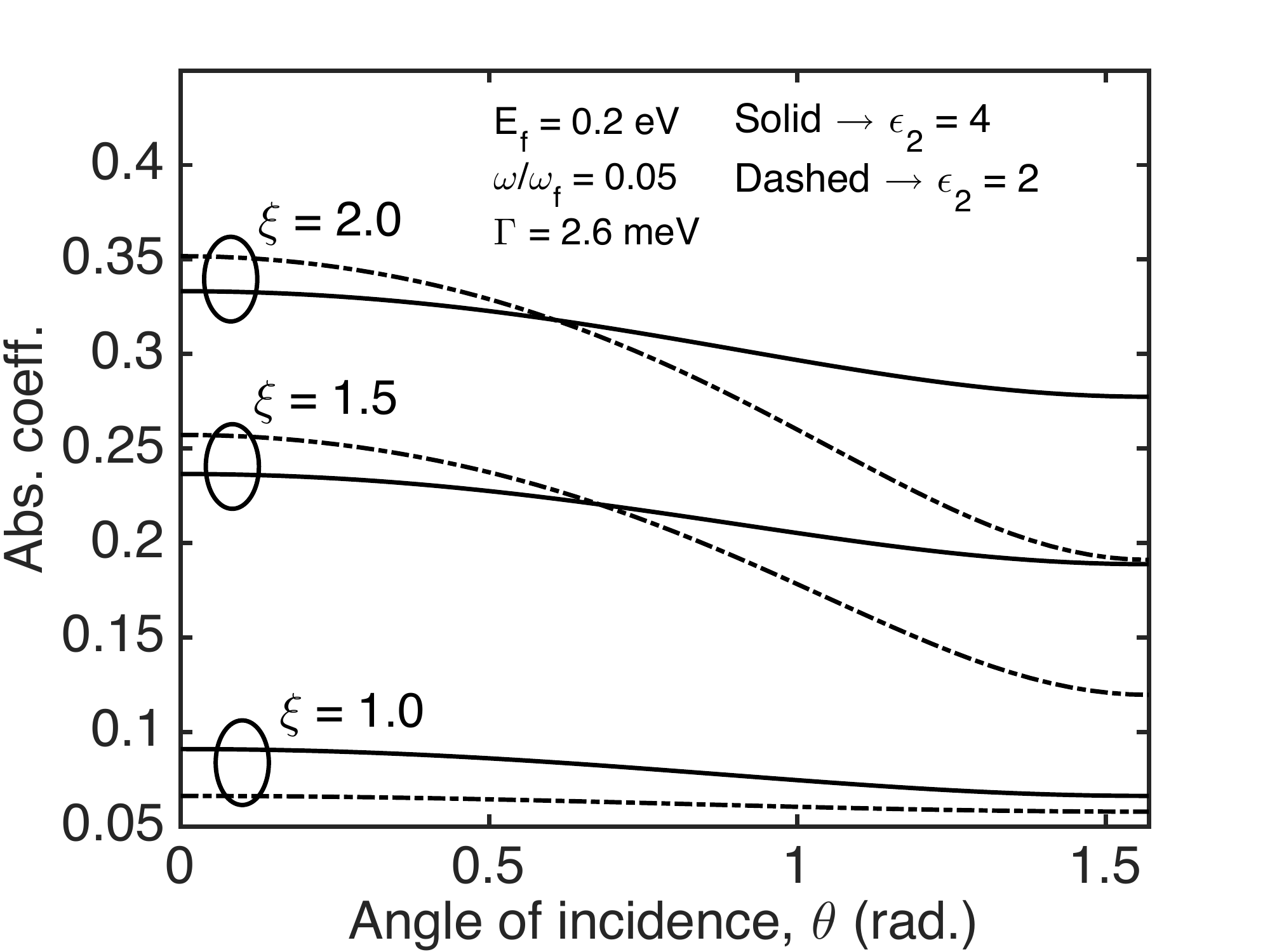}
\caption{Absorption coefficient is shown as a function of the incident angle. Different traces correspond to the marked ratio $\xi = \epsilon_{1}/\epsilon_{2}$ of the dielectric constants. Other simulation parameters are noted in the figure legend.} 
\label{fig3}
\end{figure}

\vspace{0.25cm}
The amount of light absorbed by a suspended graphene sheet must be a tunable quantity for a wide range of applications such as photo-detectors and sensors. This tuning, for a given set of material constants, can be accomplished by selecting ``control" parameters such as frequency of the incoming light-beam, the Fermi level and the angle of incidence to alter the overall reflectance, transmission, and absorption. We show in Fig.~\ref{fig3} the relation between incident angle, $\theta$, and the absorption coefficient for various values of $ \xi = \epsilon_{1}/\epsilon_{2} $ at fixed values of $E_f$, $\Gamma$, and $\omega$. The two dielectric constants surrounding the graphene sheet are denoted by $ \epsilon_{1} $ and $ \epsilon_{2} $ as sketched in Fig.~\ref{figs}. It is easily seen that the absorption coefficient reaches up to 35.0 $\%$ when the ratio of dielectric constants is 2.0 at close to normal incidence. Besides the enhanced absorption, another noteworthy feature in Fig.~\ref{fig3} is the degradation in absorption coefficient as the incident angle increases from zero to $\pi/2$. This is explained by examining the expressions for reflectance and transmittance given in Eqs.~\ref{trel2},~\ref{rel2}; for a fixed ratio of the dielectric constants, at higher angles of incidence, the transmitted wave vector (in medium ``2") $ k_{2,z} $ changes such that a greater portion of the incident light is reflected. The transmitted wave in medium ``2" is simply expressed as $ k_{2,z} = \sqrt{\left(\dfrac{\omega}{c}\right) ^{2}\epsilon_{2} - k_{x}^{2}} $, where $ k_{x} = k_{0}\sin\theta $. $ k_{0} $ and $ \omega $ describe the incident wave-vector and frequency, respectively. A lesser absorption with decreasing ratio of the dielectric constants $ \dfrac{\epsilon_{1}}{\epsilon_{2}}$ is also explained by utilizing Eqs.~\ref{trel2},~\ref{rel2}; it is evident that a lower ratio as marked on the plot will augment the reflection coefficient. 

\begin{figure} [h!]
\includegraphics[width=3.0in]{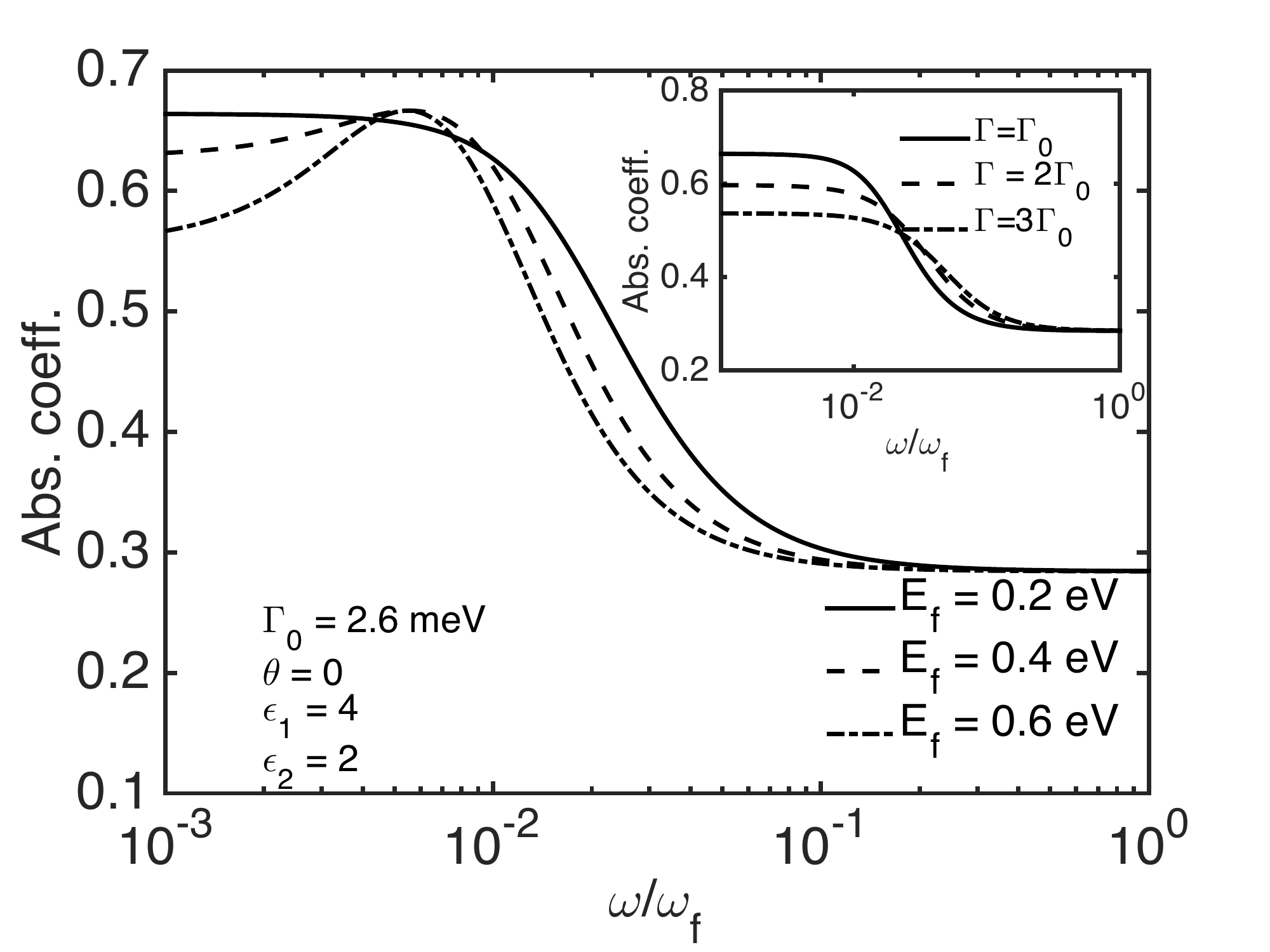}
\caption{Absorption coefficient is shown as a function of the operating frequency normalized to the frequency corresponding to the Fermi level in graphene. Different traces correspond to the marked values of Fermi level in graphene. The inset plot shows the impact of Lorentzian broadening on the absorption coefficient in graphene.} 
\label{fig4}
\end{figure}

We next turn our attention to absorption characteristics as a function of the Fermi level and the operating frequency. The functional dependence is shown in Fig. ~\ref{fig4} for absorption at normal incidence, $ \epsilon_{1} = 4 $, $ \epsilon_{2} = 2 $, and the broadening parameter $ \Gamma $ set to 2.6 $\mathrm{meV} $. At an operating frequency of around $ \omega \approx 0.01 \omega_{f} $, the absorption coefficient exhibits a peak for $ E_{f} = 0.4 \ eV $ and $ E_{f} = 0.6 \ eV $ as shown in Fig.~\ref{fig4}. Two crucial observations can be made that will help us define an ``optimal" parameter space to design graphene photonic devices. We first focus on the low $ \omega/\omega_{f} $ ratio which gives an absorption peak; at this operational frequency, the Drude inter-band conductivity dominates such that absorption increases up to approximately 70.0 $\%$, saturates, and then starts to fall. For $ \omega > 0.1 \omega_{f} $, the Drude conductivity which is still the dominant mode until $\omega = 2\omega_{f} $, rapidly drops to make a negligible contribution independent of the Fermi level(see Fig.~\ref{fig2}); with all other parameters held constant, we are therefore able to explain the merging of the absorption profiles for all three Fermi levels. The inter-band scattering is zero since we increase $ \omega $ until it reaches $ \omega_{f}$. 

The inset in Fig.~\ref{fig4} further shows the role of the broadening parameter in determining the absorption of light. As the broadening parameter gains strength, in regions where Drude conductivity is significant $ \left(\omega < 0.1\omega_f\right) $, the conductivity is lowered for identical operational characteristics such as Fermi level, frequency etc. The lowered conductivity explains the drop in absorption coefficient. The broadening parameter in the plot is a Lorentzian and is assumed to be Fermi level independent.

We have thus identified a small window for the operating frequency for $ E_{f} > 0.4$ eV, which enables a maximized absorption up to approximately 70.0 $\%$ for the selected dielectric constants surrounding the graphene sheet for TM polarized incident light. A low value of broadening parameter is desirable to keep the absorption coefficient high. While in the above estimates for absorption, the broadening was an empirically chosen number, we present below a more accurate calculation that relates it to the impurity concentration and Fermi level of graphene. 

\begin{figure} [h!]
\includegraphics[width=3.0in]{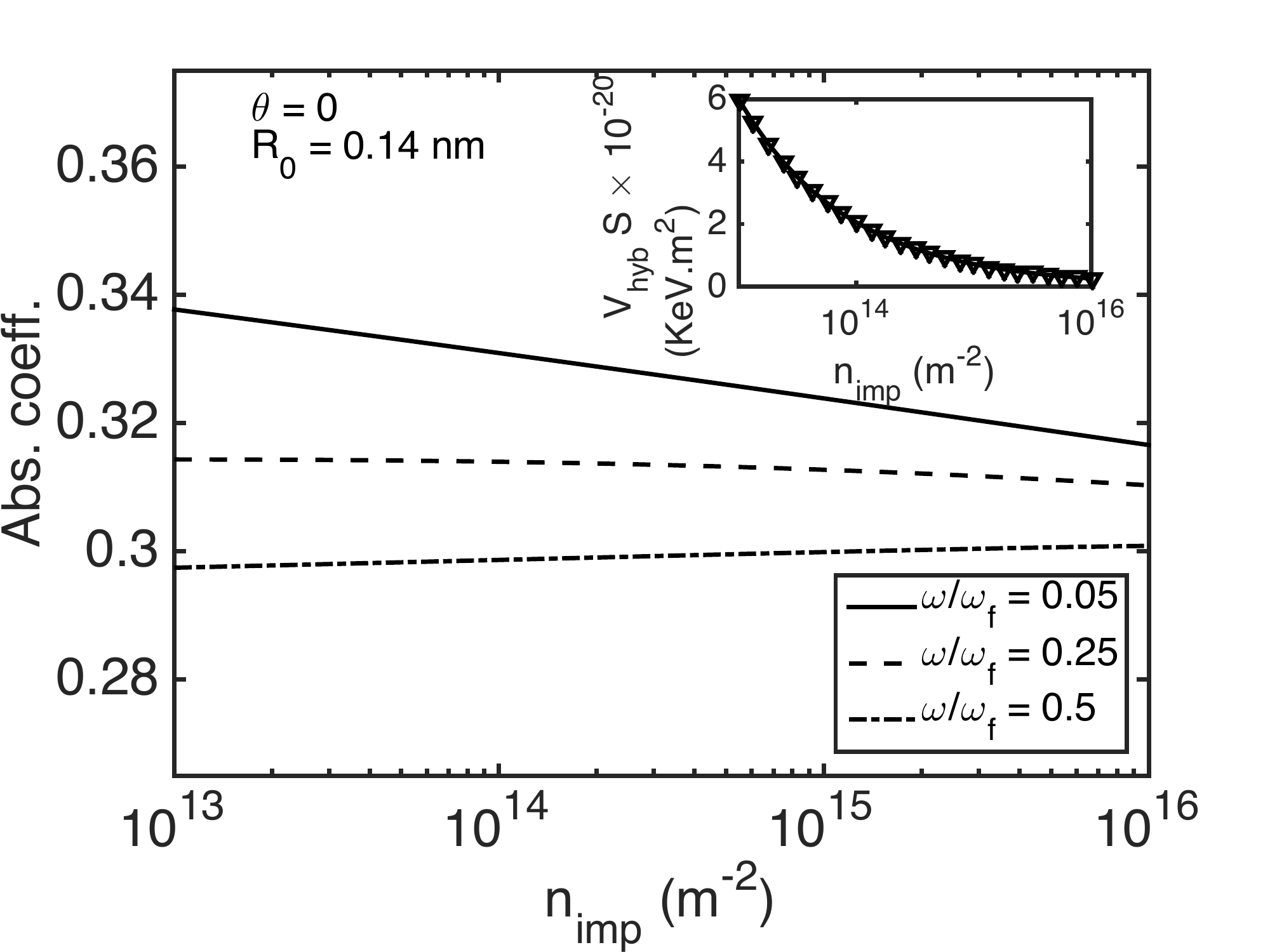}
\caption{Absorption coefficient versus impurity concentration in impure graphene for different values of operating frequency normalized to the frequency corresponding to Fermi level. The inset plot shows the hybridization potential modeled as a function of impurity concentration.} 
\label{fig5}
\end{figure}

An impure graphene sheet, which is a more realistic scenario, is considered to evaluate the spectral broadening of states. The impurities are non-interacting as described by the Hamiltonian in Eq.~\ref{hamhubb}. Figure \ref{fig5} conveys the role of impurities in influencing the optical absorption, where the Lorentzian broadening is characterized by the hybridization potential, $V_{hyb}$, of the impurity atom with the $ \pi$-bonds of graphene. The trend in Fig.~\ref{fig5} is in conformity with the behaviour shown in the inset plot of Fig.~\ref{fig4}. As expected, with an increase in the impurity concentration, the broadening is enhanced which lowers the absorption. The three trend lines shown in Fig.~\ref{fig5} further demonstrate that for a low $ \omega/\omega_{f} $ ratio, the Drude conductivity is significant as indicated by the downward trend in absorption($ \omega/\omega_{f} = 0.05 $). As $ \omega \rightarrow \omega_{f} $, Drude conductivity is considerably decreased in magnitude as depicted by the ``relatively" straight absorption curves. These absorption curves are also insensitive to the Fermi level in graphene. However, in the presence of other impurity atoms that are interacting such as charged impurities and polar phonons, absorption coefficient will indeed exhibit a dependence on the Fermi level. This aspect has not been considered. The hybridization potential used in determining the broadening is shown in the inset plot of Fig. \ref{fig5} as a function of $n_{imp}$. For a typical impurity concentration of $n_{imp} \approx 10^{10}$ $cm^{-2}$, $V_{hyb} \approx (0.5-1)$ KeV$\AA^{2}$.~\cite{adam2009theory}

We have therefore identified a few key guidelines to design a graphene photonic device. Primarily, in the low-frequency regime where Drude conductivity is active, absorption saturates as a function of the ratio of dielectric constants and frequency of incident light. The corresponding Fermi level can be adjusted such that absorption is maximized. Finally, the broadening parameter must not be too high, a large spectral broadening at low frequencies degrades the overall absorption.

\vspace{0.25cm}
\subsection{Circular Dichroism}
\begin{figure} [h!]
\includegraphics[width=3.0in]{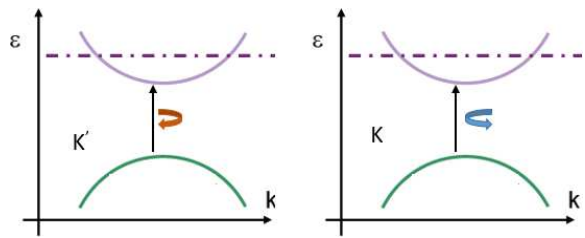}
\caption{Right and left circularly polarized light couples selectively to band edges $ K $ and $ K^{'} $ since they are time-reversed pairs. At other points in $ k $-space, not far from the band edge, the selective coupling is lost and a varying degree of polarization-dependent absorption occurs that gives rise to the phenomenon of circular dichroism. The two curved arrows denote left and right circular polarization.} 
\label{CD}
\end{figure}

\begin{figure} [h!]
\includegraphics[width=3.0in]{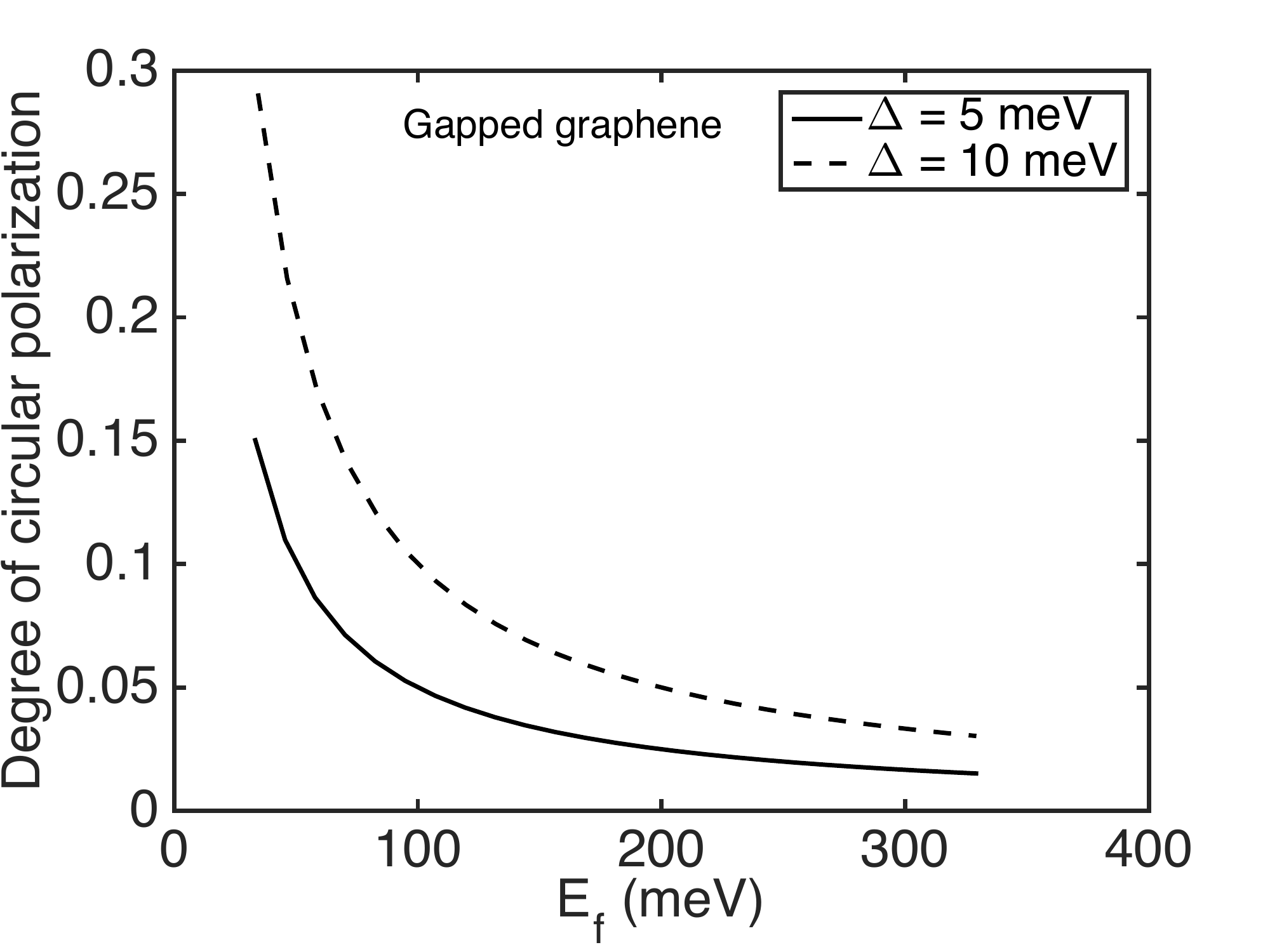}
\caption{Degree of circular polarization $ \rho $ in gapped graphene versus the Fermi level for two values of the band-gap: $\Delta$ = 5 meV and 10 meV. $ \rho $ is plotted using Eq.~\ref{avgrho} over a constant energy surface defined by the respective Fermi level, which is defined by the energy spectrum of graphene. The $ k $-vector is taken as 0.1 1/\AA. The dichroism decreases with an increasing Fermi level.} 
\label{fig6}
\end{figure}

\vspace{0.25cm}
Circular dichroism or the degree of circular polarization relates to the differential absorption of right and left circularly-polarized light. In the case of graphene with its linear bands, the dichroism is analytically computed through Eq.~\ref{dichro}. As is clear from Eq.~\ref{dichro}, pristine graphene $\left(\Delta = 0\right)$ does not exhibit circular polarization $ \rho $; however, in the presence of an inversion symmetry-breaking band gap, which renders the two sub-lattices inequivalent, $ \rho $ is a non-zero number. We seek to evaluate $ \rho $ in the following section under a symmetry-breaking condition. The coupling of right (left) circularly polarized light (see Fig.~\ref{CD}) which is exact at the Dirac point formed at $ K\left(K^{'}\right)$ edge of the Brillouin zone~\cite{yao2008valley} exhibits a polarization-dependent light absorption near a particular edge, say $ K $. This differential absorption quantified as the degree of circular polarization is plotted against the Fermi level. The Fermi level in this case is assumed to coincide with $ \sqrt{\left( \hbar v_{f}k\right)^{2} + \Delta^{2}} $. In principle, though the degree of circular polarization is a function of the momentum vector, the variation of $ \rho $  is shown for a single $ k $ point in Fig.~\ref{fig6}. 

\begin{figure} [h!]
\includegraphics[scale=0.3]{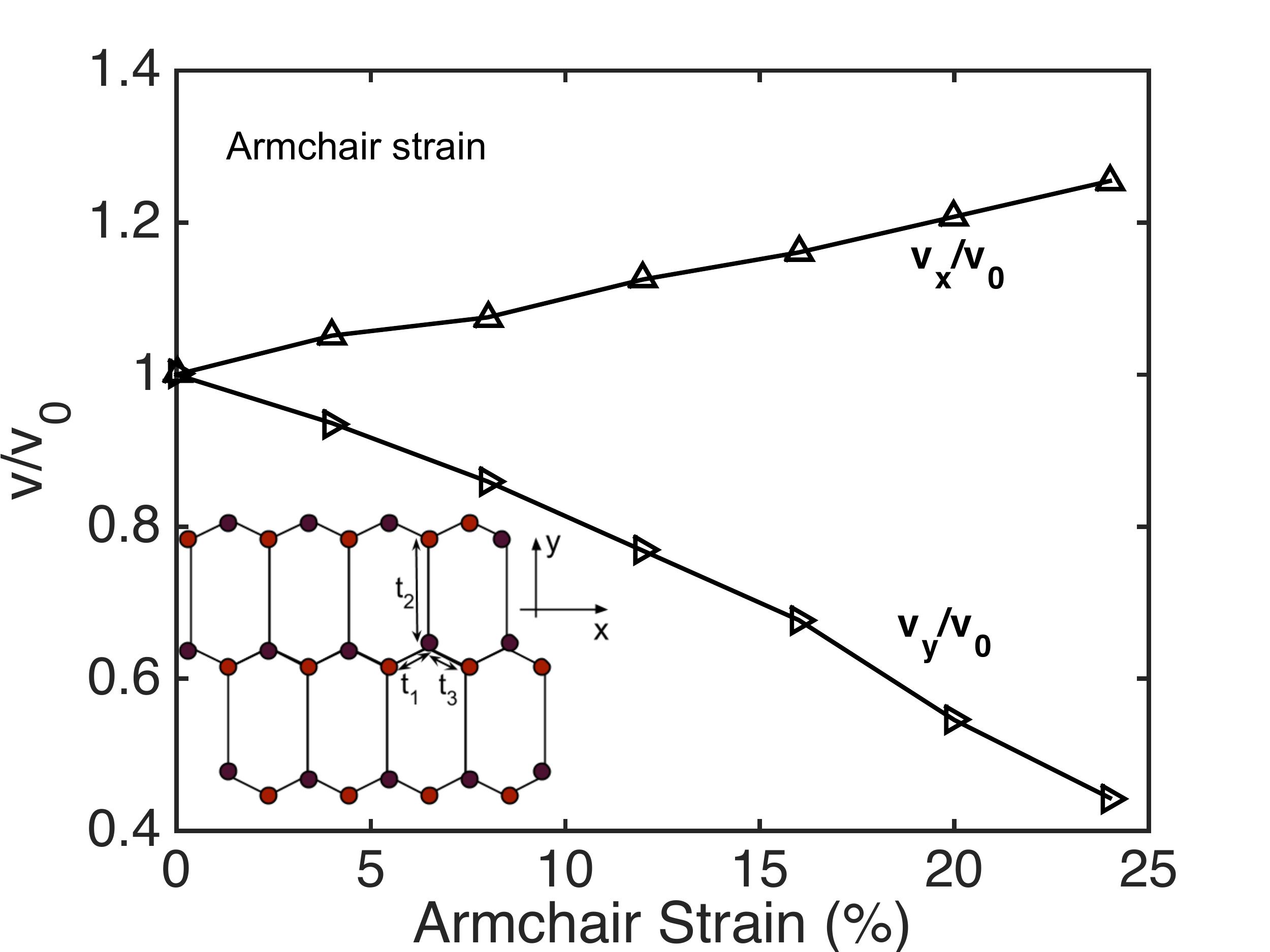} 
\caption{Data on velocities $v_y$ and $v_x$ versus uniaxial ``A'' strain as extracted from calculations in Ref.~\onlinecite{choi2010effects}. The inset shows the honeycomb lattice of graphene under strain, where the kinetic energy hopping integrals $t_i$ are shown. In this case, $t_1 = t_2 > t_3$. The inset schematically depicts distortion of the graphene honeycomb lattice under uniaxial armchair strain.}  
\label{fig7}
\end{figure}

\begin{figure} [h!]
\includegraphics[width=3.0in]{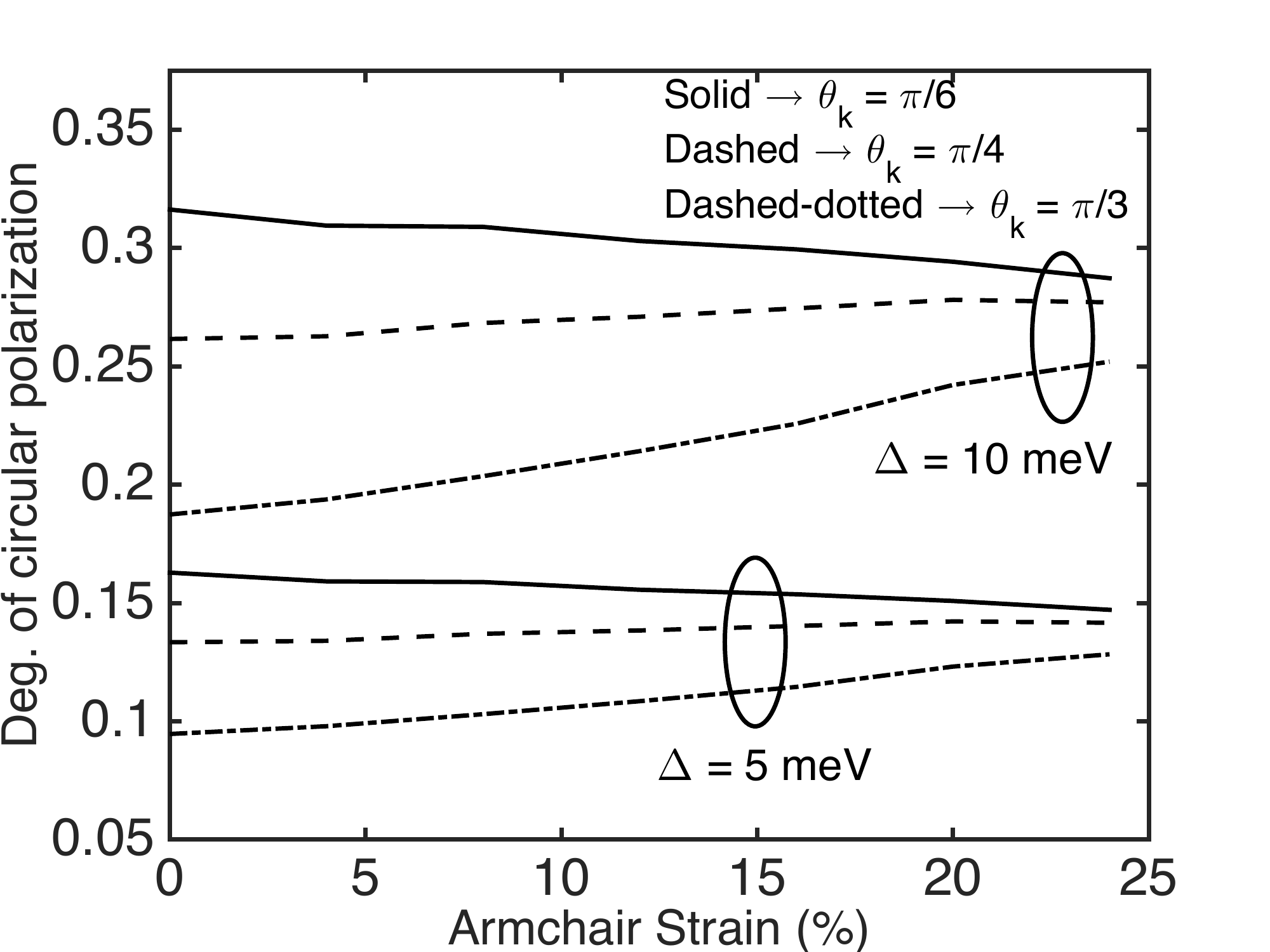} 
\caption{Degree of circular polarization in graphene versus the uniaxial armchair strain for two different values of bandgap: $\Delta$ = 5 meV and 10 meV. The degree of circular polarization depends on the anisotropy of $ k $ space and Fermi velocity components. For a larger band gap, the differential absorption of light is significant as borne out by the two well-separated group of lines for $ \Delta = 5, 10 \ meV $.} 
\label{fig8}
\end{figure}

There are several instances where graphene is not only band gap open $\left(\Delta \neq 0\right) $ at $ K $ and $ K^{'} $ edges but also asymmetrically strained (uniaxial strain) to produce tilted Dirac cones. To understand circular dichroism in this situation, we first plot in Fig.~\ref{fig7} using data from Ref.~\onlinecite{choi2010effects}, the anisotropy of $x-$ and $y-$ directed components of Fermi velocity of the Dirac fermions in a uniaxially ``A" strained graphene. The marked difference in the velocity components is evident, with an increase in ``A'' strain, the \textit{y}-component of the Fermi velocity decreases, while the \textit{x}-component is boosted such that the velocity-anisotropy factor $\kappa = v_y/v_x$ reduces to 0.35 from unity. The increase in uniaxial strain until 24$\%$ does not contribute to an additional band gap other than that created through substrate-induced optical strain. Choi \textit{et al}. in Ref.~\onlinecite{choi2010effects} explain this reduction in $ \kappa $ by considering the strength of the hopping integral parameters. The qualitative relationship between hopping integrals, connecting the nearest neighbors in a single-orbital tight-binding approximation under ``A'' strain in graphene, $t_1$ = $t_3$ $>$ $t_2$ is shown in the inset of Fig. \ref{fig7}. Using the velocity-anisotropy graph of Fig. \ref{fig7}, the degree of circular polarization in uniaxially-strained graphene is computed (Fig. \ref{fig8}) as a function of ``A'' strain percentage. We immediately notice from the plot that for $ \theta_{k} = \pi/4 $, where $ \theta_{k} = \arctan\dfrac{k_{y}}{k_{x}} $, the degree of circular polarization does not considerably vary while there is a noticeable change for $ \theta = \pi/6, \pi/3 $ with increasing strain. The trend for $ \theta = \pi/4 $ is explainable if we note that at this angle, the two components of the $ k $-vectors are symmetric, the uniaxial strain does not contribute to the band gap $ \Delta $ and the velocity anisotropy-factor only causes a slight change in degree of circular polarization. The slight deviation as strain increases and consequently the velocity-anisotropy factor $ \kappa $ supports this reasoning. The other angles introduce additional asymmetry that augments the velocity-anisotropy to produce a larger change in the degree of circular polarization. 

\begin{figure} [h!]
\includegraphics[width=3.0in]{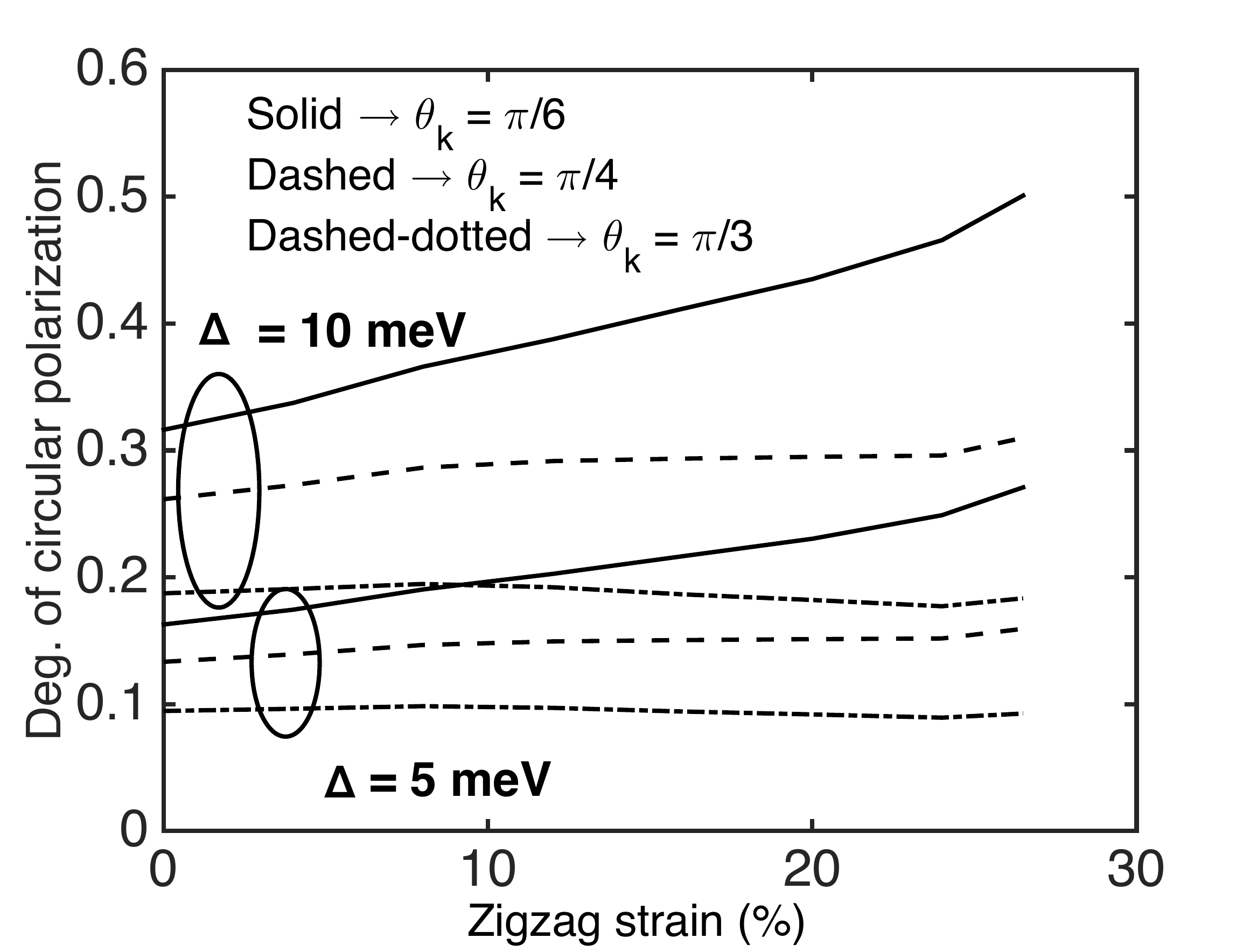} 
\caption{Degree of circular polarization in graphene versus the uniaxial zigzag strain for two different values of bandgap: $\Delta = 5 \ meV $ and $\Delta = 10 \ meV $. For $ \theta = \pi/3 $, the degree of circular polarization does not show any variation.} 
\label{fig9}
\end{figure}

The same behaviour is observed again for strain of the zig-zag type which is plotted in Fig.~\ref{fig9}. At $ \theta = \pi/4 $, the degree of circular polarization is almost constant while it shows an upward trend for $ \theta = \pi/6 $. The important point to note is the inter-play of $ k $-space and velocity anisotropy; if the two degrees of asymmetry cancel out, as it happens for $ \theta = \pi/3 $ in the zig-zag case, there will not be a sufficient variation in circular poalrization. This assumes importance in view of our proposal to enhance light absorption of a particular polarization, by selecting a suitable $ k $-space asymmetry combined with velocity anisotropy we can tune the differential absorption of right or left circularly-polarized light. This idea runs parallel to the emerging field of valleytronics in transition metal dichalcogenides.~\cite{zeng2012valley,mak2012control}
 
\vspace{0.35cm}
\section{Conclusions}
\vspace{0.35cm}
We have demonstrated in this work the tunability of optical absorption in graphene via a combination of fabrication design techniques and external dynamic control. The insertion of dielectric films of specified permitivity that surround the active graphene layer govern the transmission and reflection coefficients at a given Fermi level. The reflection and transmission coefficients which can then be made to span the complete range, $ R,T \in\left[0,1\right] $, therefore allow, through a selection of the dielectric constants, a desired absorption window. This window can be adjusted to conform to any sought level of absorption for a particular application. We refer to the choice of dielectric layers as a possible fabrication design technique. The effect of surface adsorbed impurity atoms is also a viable mechanism to tune the optical conductivity; by a careful choice of impurity atoms~\cite{chan2008first,banhart2010structural}, the spectral broadening of the density of states can be adjusted to match the proper optical absorption. Circular dichroism is considered as another possible option to selectively absorb  polarization-dependent light by focusing attention to optical processes in one of the two edges $\left(K , K^{'}\right)$. Real time dynamic control, different from fabrication designs can also be exercised by a gate contact that alters the Fermi level in the graphene sheet. A changing Fermi level influences graphene's optical conductivity and can switch the graphene-based optical device between varying degrees of absorption via modification of the reflection and transmission coefficients. 

There are several parameters that can be explored to further refine the efficiency of optical absorption in graphene. We have not considered in this work the influence of strain that changes the atomic orbital overlap to give rise to an anisotropic conductivity tensor. Strain was considered in an elementary treatment to measure the degree of circular polarization assuming that the Dirac dispersion is preserved; in reality, strain alters the density of states, the Dirac Hamiltonian and underlying optical and electronic properties. A more complete investigation of optimally engineered strained graphene structures will be considered in a follow-up work. Further, controllable doping in graphene, which provides a pathway for easy switching~\cite{giovannetti2008doping} between \textit{n}-type and \textit{p}-type can be considered as a fabrication design parameter to enhance optical absorption. Doping has not been explicitly considered in this work.

\vspace{0.35cm}
\begin{acknowledgements}
\vspace{0.35cm}
One of us (PS) expresses his gratitude to late Prof. Gabriele. F. Giuliani, Dept. of Physics, Purdue University for many illuminating discussions on optical conductivity of 2D materials that host Dirac fermions. 
\end{acknowledgements} 

\vspace{0.35cm}
\begin{appendices}
\appendix
\renewcommand \thesubsection{\Roman{subsection}}
\titlespacing\section{5pt}{12pt plus 4pt minus 2pt}{0pt plus 2pt minus 2pt}
\section{Optical absorption in mono-layer graphene}

For the sake of completeness, we derive the absorbance for a graphene sheet and show that it is independent of the incident photon frequency. The incident light is assumed to be linearly polarized along $ x $-axis and shines normally on the surface. The electric field, using the vector potential $ \overrightarrow{A}(t) = \overrightarrow{A}exp(-i\omega t) $ is
\begin{equation}
\overrightarrow{E}(t) = -\dfrac{1}{c}\dfrac{\partial \overrightarrow{A}}{\partial t}.
\label{efieldapp}
\end{equation}
The modified Hamiltonian after including the Peierls substitution takes the form $ \mathcal{H} =  v_{f}\overrightarrow{\sigma}\cdot \left(\overrightarrow{p} - \dfrac{e}{c}\overrightarrow{A}\right) $. The interaction part of the Hamiltonian is therefore $ \mathcal{H}_{int} = -\dfrac{v_{f}e}{2c}\overrightarrow{\sigma}\cdot \overrightarrow{A} $.
 
\noindent Substituting for $ \overrightarrow{A} $ from Eq.~\ref{efieldapp}, $ \mathcal{H}_{int} = \dfrac{ie v_{f}}{2\omega}\overrightarrow{\sigma}\cdot \overrightarrow{E} $.
The factor of 0.5 comes from by retaining only the $ (-i\omega t)$ term.

\noindent Using the Fermi golden rule, the transition probability for a carrier to be excited from the valence band to conduction band is $ \sum\limits_{k_{c},k_{v}}\dfrac{2 \pi}{\hbar}\vert\langle \Psi_{f}\vert \mathcal{H}_{int}\vert\Psi_{i}\rangle\vert^{2}\delta\left(E_{c} - E_{v} - \hbar\omega\right) $ where the delta function is transformed in to the density of states. 

\noindent In the above expressions, $ \Psi_{i} $ and $ \Psi_{f} $ represent the initial and final wave functions while $ E_{c} $ and $ E_{v} $ are the energies corresponding to the bottom of the conduction band and top of the valence band, respectively. Calculating the matrix element $ M\left(k\right) = \langle\Psi_{f}\vert \mathcal{H}_{int}\vert\Psi_{i}\rangle $ gives
\begin{eqnarray}
M\left(k\right) = \dfrac{1}{2}\begin{pmatrix}
\exp(i\theta) & 1
\end{pmatrix} H_{int}\begin{pmatrix}
\exp(-i\theta) \\
-1
\end{pmatrix} .
\label{fgrp}
\end{eqnarray}

\noindent Inserting the expression for $ \mathcal{H}_{int} $ in Eq.~\ref{fgrp} and noting that the $ \sigma_{x} $ dots with the $ x $ polarized electric field, one obtains
\begin{flalign}
M\left(k\right) &= \dfrac{v_{f}e\vert E_{x}\vert}{4i\omega} \begin{pmatrix}
\exp(i\theta) & 1
\end{pmatrix}\begin{pmatrix}
0 & 1 \\
1 & 0
\end{pmatrix}\begin{pmatrix}
\exp(-i\theta) \\
-1
\end{pmatrix} , \notag \\
&= \dfrac{v_{f}e\vert E_{x}\vert}{2\omega}\sin\theta, 
\end{flalign}
where the wave functions of Eq.~\ref{wfun1} are utilized. The probability density by inserting the square of the matrix element in Fermi Golden rule can be written as
\begin{equation}
P_{h\rightarrow e} = \dfrac{2\pi}{\hbar}\dfrac{v_{f}^{2}e^{2}\vert E^{2}\vert}{4\omega^{2}}\sin^{2}\theta\int d^{2}r.
\label{fgrden}
\end{equation}
The absorbed energy flux density is therefore $ \phi = \dfrac{1}{\mathcal{A}}\hbar \omega P_{h\rightarrow e}$. Summing over the density of states which is $ \dfrac{2\epsilon}{\pi \hbar^{2}v_{f}^{2}} $, where energy $ \epsilon = \dfrac{\hbar \omega}{2} $ and using the mean value of $ \sin\theta $ over a full cycle as 1/2, the final energy flux takes the following expression
\begin{flalign}
\phi_{total}&= \dfrac{1}{\mathcal{A}}\hbar \omega\dfrac{2\pi}{\hbar}\dfrac{v_{f}^{2}e^{2}\vert E^{2}\vert}{4\omega^{2}}\dfrac{1}{2}\dfrac{\hbar \omega}{2 \pi \hbar^{2}v_{f}^{2}} \\
&= \dfrac{e^{2}E^{2}}{4\hbar}.
\end{flalign}
Note that the integral $ \int d^{2}r $ is equal to $ \mathcal{A} $, the exposed graphene area in Eq.~\ref{fgrden}. The incident flux for an electric field $ \vert E \vert $ is $ \dfrac{c}{4 \pi}\vert E\vert^{2} $. The absorption is therefore $ \dfrac{\pi e^{2}}{\hbar c} = 2.3 \% $. This number includes the valley degeneracy of two in graphene.

\section{Evaluation of the commutator in Eq.~\ref{gr2}}

The final expression for the commutator in Eq.~\ref{gr2} is derived here. The commutator is given as $ \lbrace\left[ \mathcal{H}, c_{s}\left(t\right)\right] ,  c_{s}^{\dagger}\left(0\right)\rbrace $. We proceed by simplifying the inner commutator $ \left[ H, c_{s}\left(t\right)\right] $. The Hamiltonian $ \mathcal{H} $ is given by
\begin{flalign}
\begin{split}
\mathcal{H} =\sum\limits_{k}\varepsilon_{gr}a^{\dagger}_{gr}a_{gr} + \sum\limits_{s}\varepsilon_{ad}c^{\dagger}_{s}c_{s} + \sum\limits_{k}V_{hyb}a^{\dagger}_{gr}c_{s}  \\
+ \sum\limits_{k}V_{hyb}^{*}c^{\dagger}_{s}a_{gr} .
\label{hamhubbapp}
\end{split}
\end{flalign}
This $ \mathcal{H} $ gives four commutators. The first commutator is 
\begin{flalign}
\sum\limits_{s^{'}}\left[a^{\dagger}_{s^{'}}a_{s^{'}}, c_{s}\right] = \sum\limits_{s^{'}}\left(a^{\dagger}_{s^{'}}\left\lbrace a_{s^{'}},c_{s}\right\rbrace 
- \left\lbrace a^{\dagger}_{s^{'}},c_{s}\right\rbrace a_{s^{'}}\right). 
\label{comap1}
\end{flalign}
Expanding terms within the bracket, the commutator is zero. The relation $ \left[AB,C\right] =  A\left\lbrace B,C\right\rbrace - \left\lbrace A,C\right\rbrace B $ is used here. In Eq.~\ref{comap1}, the creation (annihilation) operators for graphene and the adsorbed atom are simply denoted by $ a^{\dagger}_{s}\left(a_{s}\right) $ and $ c^{\dagger}_{s}\left(c_{s}\right) $ respectively. The subscript $ s $ describes the spin projection. More detailed subscripts, on-site energies of graphene and adsorbed impurity atom, and hybridization potentials that appear in the original Hamiltonian (Eq.~\ref{hamhubbapp}) are omitted for brevity. 

\noindent The second commutator is 
\begin{flalign}
\sum\limits_{s^{'}}\left[c^{\dagger}_{s^{'}}c_{s^{'}}, c_{s}\right] = \sum\limits_{s^{'}}\left(c^{\dagger}_{s^{'}}\left\lbrace c_{s^{'}},c_{s}\right\rbrace -\left\lbrace c^{\dagger}_{s^{'}}, c_{s}\right\rbrace c_{s^{'}}\right). 
\label{comap2}
\end{flalign}
Standard commutation relations in Eq.~\ref{commrel} simplify the commutator $ \sum\limits_{s^{'}}\left[c^{\dagger}_{s^{'}}c_{s^{'}}, c_{s}\right] $ to $ -c_{s} $ .
\begin{eqnarray}
\left\lbrace c_{s^{'}},c_{s}\right\rbrace = 0 ; 
\left\lbrace c^{\dagger}_{s},c_{s}\right\rbrace = 1. 
\label{commrel}
\end{eqnarray}

\noindent The third commutator from the Hamiltonian is
\begin{flalign}
\sum\limits_{s^{'}}\left[a^{\dagger}_{s^{'}}c_{s^{'}}, c_{s}\right] = \sum\limits_{s^{'}}\left(a^{\dagger}_{s^{'}}\left\lbrace c_{s^{'}},c_{s}\right\rbrace - \left\lbrace a_{s^{'}},c_{s}\right\rbrace\right).  
\label{comap3}
\end{flalign}
Evaluating each sub-anticommutator on RHS and using commutation relations in Eq.~\ref{commrel} yields zero.

\noindent The last commutator is
\begin{flalign}
\sum\limits_{s^{'}}\left[c^{\dagger}_{s^{'}}a_{s^{'}}, c_{s}\right] = \sum\limits_{s^{'}}\left(c^{\dagger}_{s^{'}}\left\lbrace a_{s^{'}},c_{s}\right\rbrace - \left\lbrace c_{s^{'}},c_{s^{'}}\right\rbrace\right).  
\label{comap4}
\end{flalign}
As before, the evaluating the sub-anticommutators give $ \sum\limits_{s^{'}}\left[c^{\dagger}_{s^{'}}a_{s^{'}}, c_{s}\right] = -a_{s^{'}} $.

\end{appendices}

\end{document}